\documentclass[aps,twocolumn,showpacs,amsmath,letter]{revtex4}
\usepackage{graphicx, verbatim}
\newcommand{\Journal}[4]{#1 \textbf{#2}, #3 (#4)}
\usepackage{epstopdf}
\usepackage{CJK}
\usepackage[utf8]{inputenc}
\usepackage[english]{babel}
\usepackage{hyperref}
\hypersetup{colorlinks=true,linkcolor=blue,filecolor=magenta,urlcolor=cyan}

\begin{document}
\begin{CJK*}{GB}{gbsn}
\title{Control of current-induced spin-orbit effects in a ferromagnetic heterostructure by electric field}
\author{R. H. Liu(ÁõÈÙ»ª)}
\author{W. L. Lim}
\author{S. Urazhdin}
\affiliation{Department of Physics, Emory University, Atlanta, GA 30322, USA.}

\begin{abstract}
We study the effects of electrostatic gating on the current-induced phenomena in ultrathin ferromagnet/heavy metal heterostructures. We utilize heterodyne detection and analysis of symmetry with respect to the direction of the magnetic field to separate electric field contributions to the magnetic anisotropy, current-induced field-like torque, and damping torque. Analysis of the electric field effects allows us to estimate the Rashba and the spin Hall contributions to the current-induced phenomena. Electrostatic gating can provide insight into the spin-orbit phenomena, and enable new functionalities in spintronic devices.
\end{abstract}


\maketitle
\end{CJK*}
The mechanisms enabling electronic control of magnetization, such the current-induced spin transfer torque (STT)~\cite{slon1,berger,tsoiprl,cornellorig}, have been recently extensively investigated for applications in spin-based electronic (spintronic) devices. Operation of the traditional multilayer STT devices requires high current densities flowing through the magnetic layers, motivating the search for new mechanisms to control magnetic configuration. Recently,
magnetization reversal~\cite{miron2,suzuki,liulq,emori} and auto-oscillation~\cite{demidov,liurh} have been demonstrated in ferromagnet/heavy metal thin film heterostructures (FH), due to a combination of the Rashba~\cite{rashba} and the spin Hall effect (SHE)~\cite{Diakonov,hirsch} caused by the spin-orbit interaction (SOI) inside materials and/or at their interfaces. Furthermore, it was shown that SOI-induced phenomena such as magnetic anisotropy can be controlled by electric field~\cite{nakamura,maruyama,nozaki,chiba,wang_chien,leix}.
If the current- and field-induced SOI effects can be combined, it may become possible to develop spintronic devices with new functionalities, for example,  simultaneous logic and memory functions.

Here, we report a study of the effects of the electric field on the current-driven phenomena in utltrathin FHs. By using heterodyne detection\cite{heterodyne} of different harmonics of the Hall voltage produced by mixing of ac current with  ac electric field, we were able to directly separate the contributions of the electric field to the magnetic anisotropy and the current-induced torques. We found that the variation of the current-induced effective field (or the field-like torque) by the electric field $E$ can reach $4.3$\% at $E=2.8$~MV/cm, and is $8$~times larger than the effect of the electric field on the magnetic anisotropy. We also show that measurements of the electric field effects can provide insight into the fundamental mechanisms of the current-induced SOI effects. Specifically, both the Rashba spin-orbit coupling at the magnetic interface~\cite{rashba,miron1,miron2,manchon,kim2,haney} and the spin Hall effect (SHE) due to the spin-orbit scattering inside materials~\cite{liulq,Diakonov,hirsch} have been suggested as the origins for the current-induced SOI effects. Despite numerous experimental studies~\cite{liulq,miron1,miron2,pi,kim,garello}, no consensus has emerged on the dominant SOI mechanism even in the most common FHs. By analyzing the effects of the electric field, we will demonstrate that the two contributions to the field-like spin-orbit torque are comparable, while SHE provides a dominant contribution to the damping-like torque in the studied FHs.

\begin{figure}[h]

\centering
\includegraphics[width=0.45\textwidth]{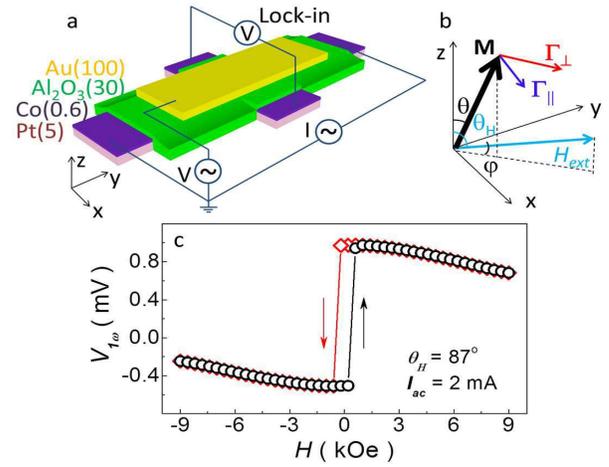}
\caption{(Color online). (a) Schematic of the device structure and experimental setup. (b) Orientations of the magnetization \textbf{M}, the external field $H_{ext}$ and the current-induced spin-orbit torques relative to the coordinate system. (c) ac Hall voltage $V_{\omega_I}$ measured at the frequency $\omega_I=23$~Hz of the ac current vs field $H$. The field is oriented at an angle $\theta_H=87^\circ$ relative to the film normal, at an in-plane angle $\varphi=90^\circ$ with respect to the direction of the ac current $I=2$~mA rms.}\label{fig1}

\end{figure}

Our device is based on a Pt(5nm)/Co(0.6nm) bilayer deposited on an annealed sapphire substrate by magnetron sputtering. The bilayer was patterned by e-beam lithography into a 4~$\mu$m-wide strip with two 2~$\mu$m-wide Hall electrodes at the sides [Fig.\ref{fig1}(a)]. A $100$~nm thick Au gate electrically isolated from the Pt/Co bilayer by a $30$~nm thick Al$_2$O$_3$ layer was fabricated on top. All measurements were performed at room temperature. The magnetic field $H$ was applied at an angle $\theta_H=87^\circ$ relative to the azimuthal direction, at an in-plane angle $\varphi=0$ or $\varphi=90^\circ$ relative to the direction of the applied ac current $I$. This orientation of the field enabled sensitive detection of the current-induced effects. We verified that the results were not significantly affected by small variations of $\theta_H$. Both the anomalous Hall effect(AHE) and the planar Hall effect (PHE) generally contribute to the Hall resistance $R_H=R/I$ of FHs. However, in the studied range of fields the magnetization of our samples was nearly normal to the film plane due to the strong perpendicular magnetic anisotropy (PMA) of the Pt(5nm)/Co(0.6nm) bilayer, and therefore the Hall voltage $V$ was dominated by the AHE.

Measurements performed at zero gate voltage $V_g=0$ allowed us to characterize the magnetic properties of the samples. The Hall resistance $R_{AH}$ characterizing the magnetic configuration was obtained from the Hall voltage at frequency $\omega=\omega_I$. The hysteresis loop obtained by scanning $H$ in the transverse configuration is consistent with a strong PMA of the Pt/Co bilayer [Fig.~\ref{fig1}(c)]. The current $I$ affected the magnetization due to a combination of the current-dependent interfacial Rashba SOI and the SHE in Pt. The effect of the Oersted field was negligible, as shown by the studies of Pt/Co bilayers with no electric gate~\cite{pi,kim,garello}. To define the framework for the analysis of the current-induced torques~\cite{manchon, haney}, we separate the damping-like contribution $\Gamma_\parallel =k_L {\mathbf m} \times [{\mathbf m}\times (\hat{z} \times {\mathbf j} )]$ from the field-like contribution  $\Gamma_{\perp} = - k_T {\mathbf m} \times (\hat{z} \times {\mathbf j})$, where the coefficients $k_L$, $k_T$ characterize the efficiency of the current-induced effects. The torques are equivalent to the effective fields
\begin{equation}\label{eqHLT}
\mathbf{H}_{SL} = - k_L \mathbf{m} \times (\hat{z} \times \mathbf{j}),  \quad \mathbf{H}_{ST}=k_T \hat{z} \times \mathbf{j}.
\end{equation}
which are more convenient for the analysis of quasiequilibrium magnetization configurations. Both $\mathbf{H}_{SL}$ and $\mathbf{H}_{ST}$ were determined as follows~\cite{pi,garello}. For the longitudinal field configuration, the orientation of the magnetization is also longitudinal, $\varphi_M=0$. The transverse effective field $H_{ST}$ then induces an in-plane rotation of ${\mathbf M}$ that does not affect the anomalous Hall resistance $R_{AH}$, while the current-induced longitudinal effective field $H_{SL}$ modulates the magnitude of the external field, resulting in a variation $\Delta R_{AH}=\frac{dR_{AH}}{dH}\frac{H_{SL}}{\sin(\theta_H-\theta)}$. An oscillating current $I_0\cos\omega_It$ produces the first and the second Hall voltage harmonics
\begin{equation}\label{eqV2w}
\begin{split}
V_{\omega_I}=I_0R_0\cos\theta \simeq\frac{I_0R_0D}{\sqrt{D^2+H^2}}\\
V_{2\omega_I}=-\frac{I_0R_0H_{SL}}{2\sin(\theta_H-\theta)}\frac{d\cos\theta}{dH} \simeq \frac{I_0R_0H_{SL}H}{2(D^2+H^2)},
\end{split}
\end{equation}
where 2$R_0$ is the difference between the anomalous Hall resistances in the up and down magnetized states, $\cos\theta=\frac{D}{(D^2+H^2\sin^2\theta_H)^{1/2}}$, $D\equiv2K_u/M_S-4\pi M_s + H_z$ is a parameter determined by the out-of-plane uniaxial anisotropy constant $K_u$, $M_s$ is the saturation magnetization, and $H_z = H\cos\theta_H$ is the out-of-plane field component. The second harmonic generated in the transverse field configuration is also proportional to $H_{ST}$, $V_{2\omega_I}=\frac{I_0R_0H_{ST}H}{2(D^2+H^2)}\cos\theta$. The general dependence on $\varphi$ is more complicated due to the rotation of the effective fields together with ${\mathbf M}$. However, one can determine $H_{SL}$ from the first and second harmonic measured in the longitudinal configuration, and $H_{ST}$ from the transverse configuration~\cite{garello,haney},
\begin{equation}\label{eqHSL}
 \begin {split}
 H_{SL,ST}=-2(\frac{dV_{2\omega_I}}{dH_{L,T}})/(\frac{d^2V_{\omega_I}}{dH_{L,T}^2})f(\theta),
 \end {split}
\end{equation}
where $f(\theta)$=1 for $H_{ST}$, $f(\theta)=\sin(\theta_H-\theta)$ for $H_{SL}$.

\begin{figure}[htbp]
\centering
\includegraphics[width=0.35\textwidth]{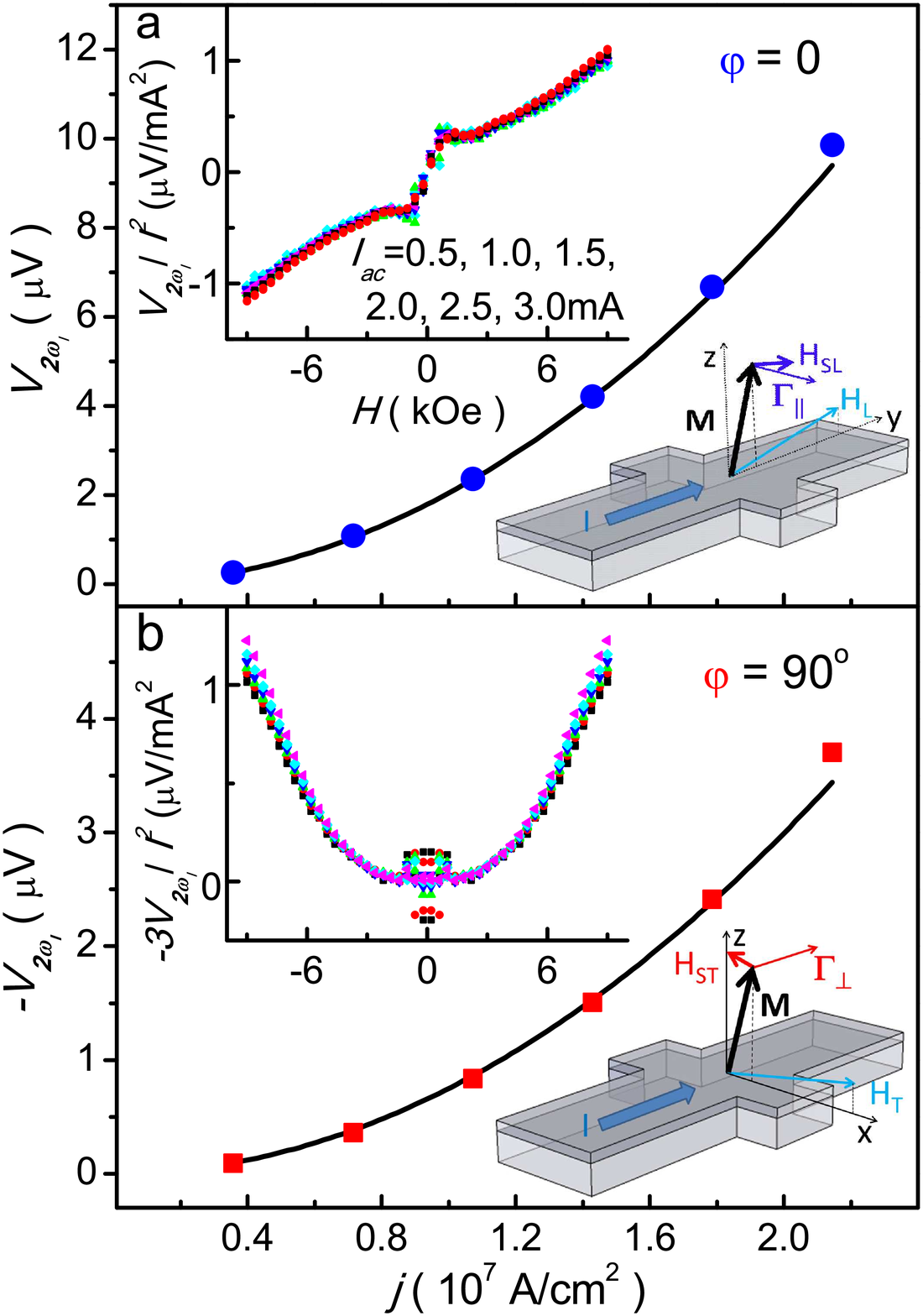}
\caption{(Color online). (a) The second Hall voltage harmonic $V_{2\omega_I}$ {\it vs} rms current density $j$ at $\varphi=0$, $H_L$=9~kOe. The top inset shows $V_{2\omega_I}/I^2$ vs $H$, at the labeled rms values of current $I$, the bottom inset is a schematic of the geometry and the torques. (b) Same as (a), at $\varphi=90^\circ$.}\label{fig2}
\end{figure}

Our measurements of $V_{2\omega_I}$ [Fig.~\ref{fig2}] are consistent with the analysis given above. In particular, $V_{2\omega_I}$ is antisymmetric with respect to $H$ at $\varphi=0$, and symmetric at $\varphi=90^\circ$, consistent with the expected symmetries of the spin torque~\cite{slon1} and the Rashba field, respectively~\cite{manchon,haney,miron2}. Furthermore, according to Eq.~(\ref{eqV2w}), $V_{2\omega_I}\propto I^2$.  This dependence is confirmed by the solid line fits in the main panels of Fig.~\ref{fig2}. Scaling by $I^2$ collapses the curves acquired at different current amplitudes [insets in Fig.~\ref{fig2}]. Minor deviations observed at large currents can be attributed to the Joule heating. By using Eq.~(\ref{eqHSL}) with $\frac{dR_{AH}}{dH}$ determined from the $V_{\omega_I}$ \textit{vs} $H$ data shown in Fig.~\ref{fig1}(c), we obtained $H_{SL} \simeq 70$~Oe at $j = 1\times 10^7$~A/cm$^2$ independent of the orientation of $\textbf{M}$. However, the field-like torque described by $H_{ST}$ is strongly dependent on the azimuthal angle $\theta_M$ of $\textbf{M}$, due to the anisotropic spin relaxation rate in FHs with strong Rashba coupling~\cite{garello,christian}. Analysis shows that $H_{ST}\simeq 35$~Oe is half of $ H_{SL}$ at $j = 1\times 10^7$~A/cm$^2$, $H$=9~kOe~\cite{SM}.

\begin{figure}[htbp]
\centering
\includegraphics[width=0.4\textwidth]{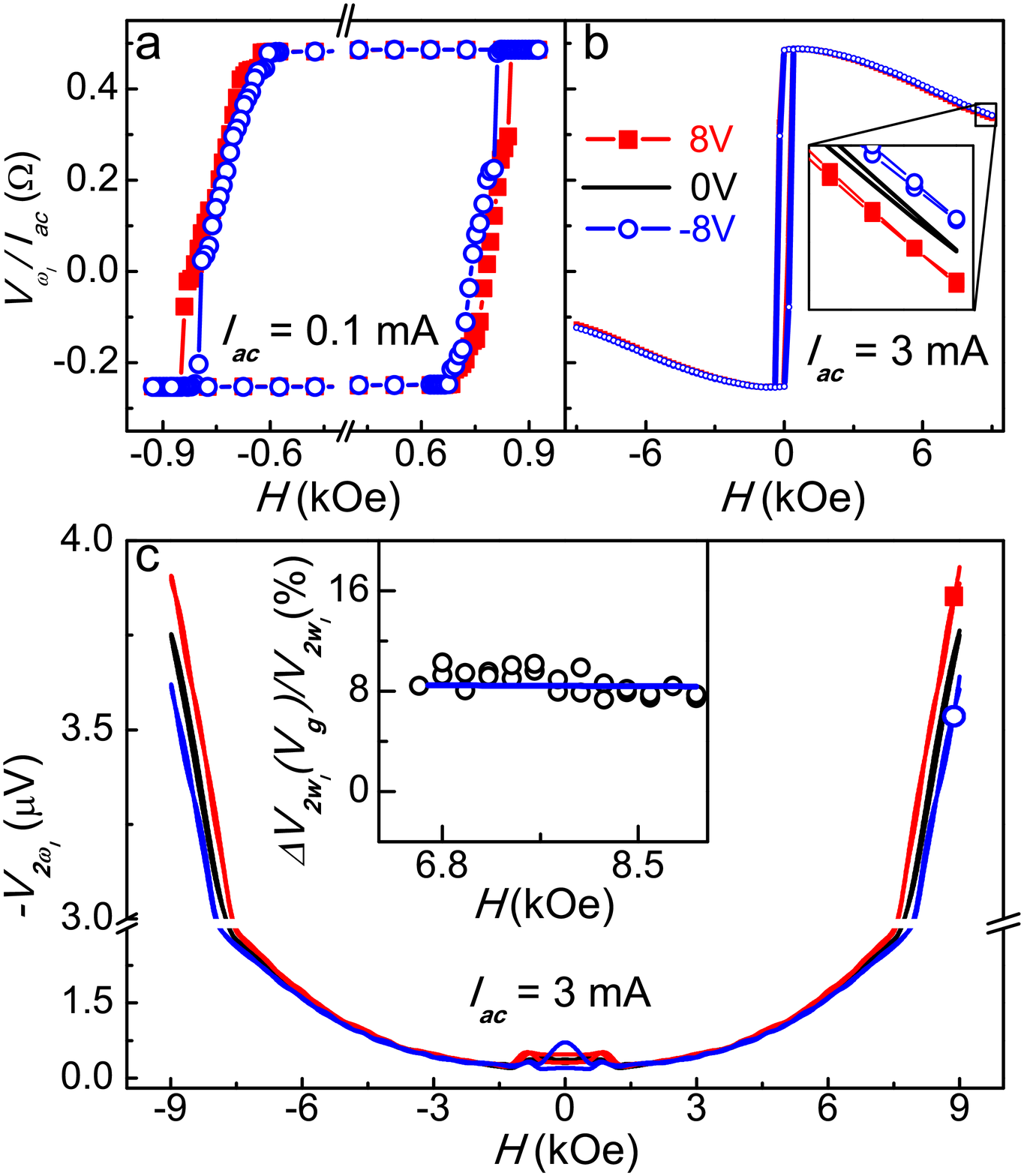}
\caption{(Color online). Effects of electrostatic gating on the current-dependent SOI. (a,b) The first harmonic Hall voltage vs transversely oriented field $H$($\varphi=90^\circ$), at gate voltage $V_g=8$~V (solid squares), $0$ (solid line), $-8$~V (open circles), at $I_{ac}=0.1$~mA in (a) and $3$~mA in (b). (c) The second harmonic Hall voltage vs $H$ measured under the same conditions as in (b). The inset is the ratio $\Delta V_{2\omega_I}/V_{2\omega_I}$, where $\Delta V_{2\omega_I}=V_{2\omega_I}(8V)-V_{2\omega_I}(-8V)$. The solid line is linear fit to date.}\label{fig3}
\end{figure}

Our central experimental result is the dependence of the current-induced effects on the electric field produced by the gate voltage $V_g$. Figure~\ref{fig3} illustrates the effects of dc electric field on the magnetic anisotropy and current-induced effective field determined from the first and the second Hall voltage harmonics. The hysteresis loop measured at the first harmonic, at $I=0.1$~mA, is broader at $V_g=8$~V than at $V_g=-8$~V [Fig.~\ref{fig3}(a)], indicating that the effective coercive field $H_c$ is enhanced at $V_g>0$. Based on the micromagnetic theory, $H_c =\alpha \frac{2K_u}{M_s} - N_{eff}M_s$, where $\alpha$ is a parameter describing the microscopic magnetic characteristics (such as magnetic pinning), and $N_{eff}$ is an effective demagnetization factor~\cite{kronmuller}. Therefore, the effect of electric field on $H_c$ comes from the modification of the magnetic pinning and/or anisotropy field~\cite{nakamura,maruyama,nozaki,chiba,wang_chien,leix}.

At $I=3$~mA, $V_{\omega_I}$ decreases with increasing $V_g$ [Fig.~\ref{fig3}(b)], while according to Eq.~(\ref{eqV2w}) it should increase due to the rotation of the magnetization towards the film normal when PMA is enhanced by $V_g$. Therefore, this variation is not only caused by the effect of gating on the anisotropy, but also by its effects on the current-induced torques. Further evidence for the effects of electric field on spin-orbit torques is provided by the dependence of the second harmonic $V_{2\omega_I}$ on $V_G$. The relative variation reached $8$\% between $V_g=8$~V and $V_g=-8$~V and is independent of $H$ [Fig.~\ref{fig3}(c)], indicating that it is dominated by the electric field-dependent current-induced torques.

\begin{figure}[htbp]
\centering
\includegraphics[width=0.4\textwidth]{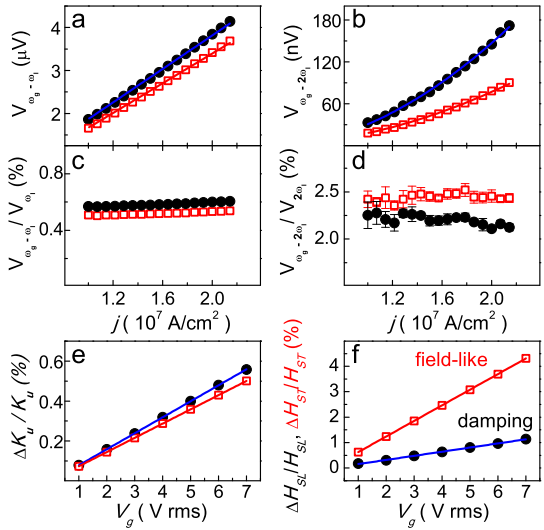}
\caption{(Color online). Quantitative analysis of the effects of electric field on SOI phenomena. (a,b) Mixing Hall voltage $V_{\omega_{g}-\omega_I}$ (a), $V_{\omega_{g}-2\omega_I}$ (b) {\it vs} current density $j$, at $V_g$=5 V rms. (c,d) $V_{\omega_{g}-\omega_I}/V_{\omega_I}$, $V_{\omega_{g}-2\omega_I}/V_{2\omega_I}$ {\it vs} $j$ determined from (a) and (b), respectively. (e) Relative variation of the magnetic anisotropy $\Delta K_u / K_u$ {\it vs} $V_g$ determined from Eq.~\ref{eqKu}. (f) Relative variation of the current-induced effective fields {\it vs} $V_g$ obtained from Eq.($\ref{eqH}$). Solid (open) symbols are data acquired at $\varphi=0$ ($\varphi=90^\circ$), solid curves are linear (a,e,f) or quadratic (b) fits to the data.
}\label{fig4}
\end{figure}

The effects of the electric field on PMA and current-induced torques can be in principle separated by analyzing the deviations of first and second Hall voltage harmonics under $V_g$ [see Eq.~\ref{eqHSL}]. We utilized an alternative approach that allowed us to directly quantify these effects. Instead of dc gate voltage, we applied an ac voltage that periodically modulated the anisotropy $K_u$ and the current-induced torques, producing an ac component of the Hall resistance at $\omega_g$. Heterodyne mixing~\cite{heterodyne} of the ac Hall resistance with ac current produced Hall voltage harmonics $V_{\omega_g\pm\omega_I}$, $V_{\omega_g\pm2\omega_I}$ at the frequencies $\omega=\omega_g\pm\omega_I$ and $\omega=\omega_{g}\pm2\omega_I$. Figure~\ref{fig4}(a,b) shows the dependence of the mixing voltages on current measured in the transverse and the longitudinal configuration. From Eq.~(\ref{eqV2w}), these voltages can be expressed as
\begin{equation}\label{eqVg}
V_{\omega_g-\omega_I} = A\Delta K_u, V_{\omega_g-2\omega_I}= B\Delta H_{SL}-C\Delta K_u
\end{equation}
where A=$\frac{2R_0I_0H^2\cos\theta}{M_s(D^3+DH^2)}$, B=$\frac{R_0I_0H}{2(D^2+H^2)}$, C=$\frac{2R_0I_0DHH_{SL}}{M_s(D^2+H^2)^2}$, $\Delta K_u$ and $\Delta H_{SL}$ are the modulations by the gate voltage of the perpendicular magnetic anisotropy constant and of the current-induced effective field, respectively. The parameter D is obtained from the dependence of $V_{\omega_I}$ on $H$ [Fig.~\ref{fig1}(c)]. According to Eq.~(\ref{eqVg}), the dependence of $V_{\omega_g\pm\omega_I}$, $V_{\omega_g\pm2\omega_I}$ on current should be the same as that of $V_{\omega_I}$ and $V_{2\omega_I}$, respectively. Indeed, their ratios $V_{\omega_g-\omega_I}/V_{\omega_I}$ and $V_{\omega_g-2\omega_I}/V_{2\omega_I}$ at a given $V_g$ are independent $I$ [Fig.~\ref{fig4}(c,d)].

To express the effect of $V_g$ on $K_u$ in terms of mixing voltages, we rewrite Eq.~(\ref{eqVg}) as
\begin{equation}\label{eqKu}
\frac{\Delta K_u}{K_u}=\frac{D(D^2+H^2)}{H^2(D+4\pi M_s)}\frac{V_{\omega_{g}-\omega_I}}{V_{\omega_I}},
\end{equation}
The modulation of $K_u$ determined from Eq.~\ref{eqKu} exhibits a linear dependence on $V_g$, with the relative variation reaching $0.55$\% for $\varphi = 0^\circ$, and $0.5$\% for $\varphi = 90^\circ$ at $V_g$ = 7 V rms with $M_s$ = 1700~emu/cm$^3$, $K_u$ = 2.6$\times$10$^7$~erg/cm$^3$ and $D = 9.5$~kOe [Fig.~\ref{fig4}(e)].

Since the heterodyne signal does not contain phase information, the signs of the field-induced effects must be separately determined. The anisotropy coefficient $K_u$ is enhanced at $V_g>0$, resulting in a decrease of $V_{2\omega_I}$ [Fig.~\ref{fig3}(a)]. Since $V_g>0$ actually increases $V_{2\omega_I}$ at $\varphi=90^\circ$ [Fig.~\ref{fig3}(c)], the contribution from the gate voltage-dependent current-induced field is opposite to that of $K_u$, and therefore
\begin{equation}\label{eqH}
\begin{split}
\frac{\Delta H_{ST}}{H_{ST}}=\frac{V_{\omega_{g}-2\omega_I}}{V_{2\omega_I}}+\frac{2D^2-H^2_T}{H^2_T}\frac{V_{\omega_{g}-\omega_I}}{V_{\omega_I}},\\
\frac{\Delta H_{SL}}{H_{SL}}=\frac{V_{\omega_{g}-2\omega_I}}{V_{2\omega_I}}-\frac{2D^2}{H^2_L}\frac{V_{\omega_{g}-\omega_I}}{V_{\omega_I}},
\end{split}
\end{equation}

Figure \ref{fig4}(f) shows the modulation of the current-induced effective fields by the gating voltage determined from the mixing Hall voltages using Eq.~(\ref{eqH}). The amplitude of the modulation is linear in $V_g$, and at $V_g=7$~V rms reaches $1.1$\% and $4.3$\% for the longitudinal and transverse field configurations, respectively. We note that the effect of the electric field on the field-like torque ($H_{ST}$) is about $8$ times larger than on the magnetic anisotropy [Fig.~\ref{fig4}(e)].

In addition to electrical control of current-induced phenomena, voltage-dependent SOI phenomena can also elucidate the mechanisms of the current-induced torques in FHs. Both the bulk SHE originating from the spin-orbit scattering inside Pt and the Rashba-type spin-orbit interaction at the Co/Pt interface and/or Co/oxide interface have been suggested as the dominant SOI mechanisms~\cite{miron2,liulq,rashba,Diakonov,hirsch,miron1,manchon,kim2,haney}. For the Pt(5)/Co(0.6) bilayer used in our experiment, the diffusion model predicts that the contribution of the Rashba SOI to the field-like torque $\Gamma_\perp$ is approximately twice its contribution to the damping-like torque $\Gamma_\parallel$, while for the SHE $\Gamma_\perp/\Gamma_\parallel\sim 0.2$~\cite{manchon2}. The latter has been also supported by previous experimental measurements~\cite{fan2}. Because the electric field is efficiently screened within Co(0.6), it is only expected to modify the interfacial Rashba contribution, while its effect on the SHE should be minimal. While the electric field at the Co/Pt interface is expected to be significantly smaller than at the AlO$_x$/Co due to screening in Co, but its contribution to the modification of the Rashba effect at the Co/Pt interface cannot be entirely eliminated, due to the local thickness variations of the Co layer. Based on the data of Fig.~\ref{fig4}(f) and assuming $\Gamma_\perp/\Gamma_\parallel= 0.2$ for SHE, $\Gamma_\perp/\Gamma_\parallel = 2$ for the Rashba SOI, the contribution of SHE to the damping-like torque is $5$ times larger than that of the interfacial Rashba effect in the studied AlO$_x$/Co/Pt system~\cite{SM}. This also shows that the current-induced
total damping-like torque is twice as large as the total field-like torque, which is consistent with the direct comparison of the damping-like and field-like torques obtained by measurements of Hall voltages at $V_g=0$~\cite{garello}.

In summary, we have demonstrated the possibility to control not only the magnetic anisotropy, but also the current-induced spin-orbit effects in magnetic heterostructures with strong spin-orbit interaction. By analyzing the effects of electric gating on the spin-orbit interaction phenomena, we demonstrated that the electric field modulating the current-induced field-like torque is $8$ times larger than the direct effect of gating on the magnetic anisotropy. Furthermore, by analyzing the symmetry of the current-induced spin-orbit torques and the  electric-field effects, we demonstrated that spin current produced by the SHE is the dominant source for current-induced phenomena in the studied system. However, the interfacial Rashba effect also provides a considerable contribution to current-induced torque. This contribution is significantly affected by the electric field, making it possible to electrically modulate the current-induced phenomena. By enhancing the interfacial Rashba SOI in optimized magnetic heterostructures, it may become possible to electrically control current-induced phenomena such as magnetization reversal. Electrical modulation of the current-induced precession can be utilized for frequency mixing or in the feedback circuits to improve the oscillation characteristics.

This work was supported by NSF DMR-1218414, ECCS-1218419, and ECCS-1305586.


\begin{references}
\bibitem{slon1}
J. C. Slonczewski. J. Magn. Magn. Mater., \textbf{159}, L1-L7 (1996); J. Magn. Magn. Mater., \textbf{195}, L261-L268 (1999).

\bibitem{berger}
L. Berger. Phys. Rev. B. \textbf{54}, 9353(1996); J. Appl. Phys., \textbf{90}, 4632 (2001).

\bibitem{tsoiprl} M. Tsoi, A.G.M. Jansen, J. Bass, W.C. Chiang, M. Seck, V. Tsoi and P. Wyder, \Journal{Phys. Rev. Lett.}{80}{4281}{1998}; \textbf{81}, 493(E) (1998).

\bibitem{cornellorig} J.A. Katine, F.J. Albert, R.A. Buhrman, E.B. Myers, and D.C. Ralph, \Journal{Phys. Rev. Lett.}{84}{3149}{2000}.

\bibitem{miron2}
I. M. Miron, K. Garello, G. Gaudin, P. J. Zermatten, M. V. Costache, S. Auffret, S. Bandiera, B. Rodmacq, A. Schuhl, and P. Gambardella. \Journal{Nature}{476}{189}{2011}.

\bibitem{suzuki}
T. Suzuki, S. Fukami, N. Ishiwata, M. Yamanouchi, S. Ikeda, N. Kasai, and H. Ohno. \Journal{Appl. Phys. Lett.} {98}{142505}{2011}.

\bibitem{liulq}
L. Q. Liu, O. J. Lee, T. J. Gudmundsen, D. C. Ralph, and R. A. Buhrman. \Journal{Phys. Rev. Lett.} {109} {096602} {2012}.

\bibitem{emori}
S. Emori, et al., \Journal{Nat. Mater.}{12}{611}{2013}.

\bibitem{demidov}
V. E. Demidov, S. Urazhdin, H. Ulrichs, V. Tiberkevich, A. Slavin, D. Baither, G. Schmitz, and S. O. Demokritov.
 Nat. Mater. \textbf{11}, 1028 (2012).

\bibitem{liurh}
R. H. Liu, W. L. Lim, and S. Urazhdin. Phys. Rev. Lett. \textbf{110}, 147601 (2013)

\bibitem{rashba}
Y. A. Bychkov and E. I. Rashba, J. Phys. C \textbf{17}, 6039 (1984).
G. Dresselhaus. Phys. Rev. \textbf{100}, 580 (1955).

\bibitem{Diakonov}
M. I. Dyakonov and V. I. Perel, \Journal{Sov. Phys. JETP Lett}{13}{467}{1971}.

\bibitem{hirsch}
J. E. Hirsch. Phys. Rev. Lett. \textbf{83}, 1834 (1999).

\bibitem{nakamura}
K. Nakamura, R. Shimabukuro, et al. \Journal{Phys. Rev. Lett.}{102}{187201} {2009}. M. Tsujikawa and T. Oda. \Journal{Phys. Rev. Lett.}{102}{247203} {2009}.

\bibitem{maruyama}
T. Maruyama, Y. Shiota, T. Nozaki, et al. \Journal{Nat. Nanotechnol.} {4} {158} {2009}.

\bibitem{nozaki}
T. Nozaki, Y. Shiota, M. Shiraishi, et al., \Journal{Appl. Phys. Lett.} {96} {022506} {2010}.

\bibitem{chiba}
D. Chiba, S. Fukami et al., \Journal{Nat. Mater.} {10} {853} {2011}.

\bibitem{wang_chien}
W. G. Wang, M. Li, S. Hageman and C. L. Chien. \Journal{Nat. Mater.} {11} {64} {2012}.

\bibitem{leix}
L. Xu and S. F. Zhang. \Journal{J. Appl. Phys.} {111} {07C501} {2012}.

\bibitem{heterodyne}
R. F. Graf, Modern Dictionary of Electronics (7th ed.,) Elsevier Inc. (1999).

\bibitem{miron1}
I. M. Miron, G. Gaudin, S. Auffret, B. Rodmacq, A. Schuhl, S. Pizzini, J. Vogel, and P. Gambardella. \Journal{Nat. Mater.} {9} {230} {2010}.

\bibitem{manchon}
A. Manchon, and S. Zhang. \Journal{Phys. Rev. B.} {79}{094422} {2009}.


\bibitem{kim2}
K. W. Kim, J. H. Moon, K. J. Lee, and H. W. Lee. \Journal{Phys. Rev. Lett.} {108} {217202}{2012}.

\bibitem{haney}
P. M. Haney, H. W. Lee, K. J. Lee, A. Manchon, and M. D. Stiles. \Journal{Phys. Rev. B.} {87} {174411}{2013}.

\bibitem{pi}
U. H. Pi, K. W. Kim, J. Y. Bae, S. C. Lee, Y. J. Cho, K. S. Kim, and S. Seo. \Journal{Appl. Phys. Lett.} {97}{162507}{2010}.

\bibitem{kim}
J. Kim, J. Sinha, M. Hayashi, M. Yamanouchi, S. Fukami, T. Suzuki, S. Mitani, and H. Ohno. \Journal{Nat. Mater.} {12}{240}{2013}.

\bibitem{garello}
K. Garello, I. M. Miron, C. O. Avci, F. Freimuth, Y. Mokrousov, S. Bl\"{u}gel, S. Auffret, O. Boulle, G. Gaudin, and P. Gambardella. \Journal{Nat. Nanotech.} {8} {587} {2013}.

\bibitem{SM}See Supplemental Material at \href{http://journals.aps.org/prb/abstract/10.1103/PhysRevB.89.220409\string#supplemental}{http://link.aps.org/supplemental/
10.1103/PhysRevB.89.220409} for more detailed analysis of
current-induced effective fields and the contributions of spin
Hall and Rashba effects.

\bibitem{christian}
C. O. Pauyac, X. H. Wang, M. Chshiev, A. Manchon. \Journal{Appl. Phys. Lett.} {102} {252403} {2013}.

\bibitem{kronmuller}
H.Kronm\"{u}ller, \Journal{Phys. Stat. Sol. (B)} {144} {385} {1987}.

\bibitem{manchon2}
A. Manchon, et al., arXiv:1204.4869 (2012).

\bibitem{fan2}
X. Fan, et al., \Journal{Nat. Comm.}{5}{3042}{2014}.


\noindent

\end{references}
\end{document}